\documentclass[aps,prd,10pt,superscriptaddress,twocolumn,shortbibliography,showkeys]{revtex4-2}
\usepackage{graphicx}
\usepackage{xcolor}
\usepackage{hyperref}
\usepackage{amsfonts, amsmath, amssymb}
\usepackage{lmodern}
\usepackage{dcolumn}
\usepackage{multirow}
\usepackage{enumitem}
\usepackage{booktabs}
\usepackage{enumitem}
\usepackage{subcaption}
\usepackage[sort&compress]{natbib}
\usepackage{physics}
\usepackage{bm}
\renewcommand{\d}{\mathrm{d}}

\hypersetup{colorlinks=true, linkcolor=blue, urlcolor=blue, citecolor=blue}

\begin{document}
\sloppy
\title{Beyond $\Lambda$CDM with a Logistic RG-like Flow of the Low Redshift Cosmic Evolution}
\author{Shibendu Gupta Choudhury}
\email{pdf.schoudhury@jmi.ac.in}
\author{Anjan A Sen}
\email{aasen@jmi.ac.in}
\affiliation{Centre for Theoretical Physics, Jamia Millia Islamia, New Delhi-110025, India}

\begin{abstract}
    Recent observations hint at possible late-time deviations from $\Lambda$CDM. We introduce a minimal phenomenological framework in which the total equation of state $w_{\rm T}(z)$ follows a logistic evolution motivated by renormalization-group–like flow between cosmological fixed points. Using DESI-DR2 BAO, DES supernova data, and CMB distance priors, we find that this parametrization provides an improved description of the expansion history relative to $\Lambda$CDM. The reconstructed evolution, though model specific, also shows statistically strong low-redshift deviations from $\Lambda$CDM.
\end{abstract}

\maketitle

\section{Introduction}
The late-time acceleration of the Universe is commonly attributed to dark energy (DE) with a constant equation of state $w_{\rm DE}=-1$, corresponding to the cosmological constant. This defines the concordance $\Lambda$CDM model, which successfully describes a wide range of observations, including the cosmic microwave background (CMB)~\cite{Planck:2018vyg,Planck:2015bue,ACT:2020gnv,ACT:2025fju}, large-scale structure~\cite{BOSS:2014hhw,BOSS:2016wmc,Addison:2017fdm,Haridasu:2017ccz}, and Type Ia supernovae (SNIa)~\cite{Riess:2019qba,Pan-STARRS1:2017jku}.

Despite its observational success, the $\Lambda$CDM paradigm faces a number of theoretical challenges~\cite{Weinberg:2000yb,Padmanabhan:2002ji}. Moreover, recent observations have started giving hints at possible deviations from this framework. In particular, baryon acoustic oscillation (BAO) measurements from the Dark Energy Spectroscopic Instrument (DESI)~\cite{DESI:2024mwx,DESI:2025zgx}, combined with SNIa~\cite{Rubin:2023ovl,DES:2024jxu,DES:2025sig,Brout:2022vxf} and CMB data~\cite{Planck:2018vyg,ACT:2020gnv}, indicate a preference for evolving DE at late times, potentially involving a phantom crossing in the past. With improved calibration of the Dark Energy Survey (DES) 5-year SNIa sample, the deviation from $\Lambda$CDM reaches a statistical significance of $\sim 3.2\sigma$, motivating alternative descriptions of cosmic acceleration~\cite{DiValentino:2021izs,Abdalla:2022yfr}.

Reconstruction techniques provide a powerful framework to probe such deviations in a relatively model-independent way~\cite{Sahni:2006pa}. 
These approaches include both direct reconstructions of the expansion history~\cite{Holsclaw:2010sk,Holsclaw:2011wi,Sahlen:2005,Huterer:2003, Mukherjee:2024pcg, Mukherjee:2024ryz, Mukherjee:2025ytj, GuptaChoudhury:2025uff} and parametrized reconstructions involving physically motivated modifications~\cite{Starobinsky:1998fr,Huterer:1998qv,Chevallier:2000qy,Sen:2001xu,Sen:2005ra,Choudhury:2025bnx}. A commonly adopted strategy is to parametrize the DE equation of state $w_{\rm DE}$~\cite{Chevallier:2000qy,Linder:2002et,Linder:2005in,Capozziello:2005ra, Jassal:2005qc, Barboza:2008rh}. However, the background expansion of the Universe depends only on the total energy density and pressure, implying that observations measuring distances directly probe the total equation of state \(w_{\rm T}\), rather than that of DE alone. Consequently, $w_{\rm DE}$ should be regarded as an effective description~\cite{Caldwell:2025inn,Liu:2025bss} requiring additional assumption about the matter density $\Omega_m(z)$ to obtain the expansion dynamics. This motivates reconstructing $w_{\rm T}$ as a more fundamental quantity characterizing the cosmic expansion history. 

In this work, we explore a novel framework in which the evolution of $w_{\rm T}$ is governed by dynamics reminiscent of a renormalization group (RG) flow. Specifically, we consider a logistic RG flow like evolution that smoothly interpolates between the matter dominated epoch and the present phase of accelerated expansion. In contrast to widely used parametrizations, such as the Chevallier-Polarski-Linder (CPL)~\cite{Chevallier:2000qy, Linder:2002et, Linder:2005in}, Jassal-Bagla-Padmanabhan~\cite{Jassal:2005qc}, Barboza-Alcaniz~\cite{Barboza:2008rh}, Scale factor parametrization~\cite{Sen:2001xu, Mukhopadhyay:2024fch} etc., the $\Lambda$CDM model is not embedded within this framework. It therefore provides an independent and minimally biased probe of deviations from the standard cosmological scenario, without being anchored to it as a limiting case. We examine whether current observations show a preference for such a logistic evolution of the cosmic expansion.\\

\section{Theoretical framework}
From a general perspective, the cosmic expansion can be interpreted as a scale evolution problem: as the Universe expands, its characteristic energy scale decreases and the effective cosmological dynamics changes. This perspective naturally suggests an analogy with RG flows in quantum field theory, where physical parameters evolve with the scale: $\bm{\d g/\d \mu = \beta(g)}$,
$g$ being the physical parameter that varies with the energy scale $\mu$ and $\beta(g)$ is the beta-function governing the flow of $g$.
Analogously, in cosmological evolution, the redshift (which is measured directly) plays the natural role of the flow parameter. As the Universe expands, the effective cosmological dynamics evolves from radiation dominated to matter dominated and finally towards the accelerating phase with specific values for $w_{\rm T}$ (which plays the role of the physical parameter $g$) at each epoch. This transition can therefore be interpreted as a flow between different dynamical fixed points of the cosmic equation of state $w_{\rm T}$.

Here, we are primarily interested in the low-redshift Universe, which interpolates between a matter-dominated phase with $w_{\rm T}\approx 0$ and a late-time accelerating epoch with $w_{\rm T}<-1/3$. Motivated by the RG flow analogy, we describe the evolution of $w_{\rm T}$ through a beta-function type equation,
\begin{equation}\label{rgeq}
\frac{d w_{\rm T}}{dz} = \beta(w_{\rm T}),
\end{equation}
where $\beta(w_{\rm T})$ characterizes the scale evolution of $w_{\rm T}$. The simplest beta function capable of generating two fixed points is quadratic, $\bm{\beta(w_{\rm T}) = \alpha w_{\rm T}(\gamma-w_{\rm T})}$,
where $\alpha$ and $\gamma$ are constants. This form admits the fixed points: $w_{\rm T}=0$ and $w_{\rm T}=\gamma$, the latter represents the late-time accelerating phase for $\gamma<-{1}/{3}$. This structure is characteristic of logistic evolution equations that commonly arises in dynamical systems and RG flows connecting two fixed points.

Logistic type evolution also appears naturally in cosmology~\cite{Scherrer:2005je, Scherrer:2007pu}. For instance, in the $\Lambda$CDM model the DE density parameter evolves as
\begin{equation}\label{dwtlcdm}
\frac{d\Omega_{\Lambda}}{d \log a}=3\,\Omega_{\Lambda}(1-\Omega_{\Lambda}) \rightarrow \frac{dw_\text{T}^{\Lambda}}{d \log a}=3\,w_\text{T}^{\Lambda}(1+w_\text{T}^{\Lambda}),
\end{equation}
which interpolates between the fixed points $\Omega_{\Lambda}=0$ and $\Omega_{\Lambda}=1$, equivalently, between the fixed points $w_\text{T}^{\Lambda}=0$ and $w_\text{T}^{\Lambda}=-1$. These are logistic equations with $\log a$ as the flow parameter instead of $z$. An important consequence of this relation concerns the jerk parameter,
$\bm{j = \frac{\dddot a}{aH^3}}$,
which can be expressed as
\begin{equation}\label{jerkwt}
j(z) = 1 + \frac{9}{2} w_T(1+w_T) + \frac{3}{2}(1+z)\frac{dw_T}{dz}.
\end{equation}
Using Eq.~\eqref{dwtlcdm} it follows that $j=1$ at all redshifts in the $\Lambda$CDM model.

If we seek a more general description, we can make the constants in Eq.~\eqref{dwtlcdm} to be arbitrary. It can be shown that such a form effectively corresponds to a cosmological scenario consisting of pressureless matter and DE with a constant equation of state, which has been widely studied in the literature.  To obtain more non-trivial evolution, we instead adopt Eq.~\eqref{rgeq} as the flow equation for $w_{\rm T}$ with a quadratic beta-function $\beta(w_{\rm T})$ as mentioned above. With a convenient redefinition of parameters, the evolution equation for $w_{\rm T}$ and a particular solution of it can be written as
\begin{equation}\label{dwtlogis}
\frac{dw_{\rm T}}{dz} = -\frac{B  w_{\rm T}\left(A + w_{\rm T}\right)}{A}
\rightarrow w_{\rm T}(z) = \frac{-A}{1+\exp(B z)}.
\end{equation}
The integration constant is fixed to obtain a minimal two-parameter form that captures the transition between fixed points while avoiding additional degeneracies. We refer to this as the Logistic model in the subsequent discussion. The resulting $w_{\rm T}$ has a sigmoid (logistic) form similar to the $\Lambda$CDM case, however, it  does not reproduce $\Lambda$CDM or CPL behaviour for any choice of parameters. At this stage, we want to stress on a few points. All the RG-flow logistic type equations mentioned above are autonomous equations in terms of the flow parameters (either $z$ or $\log a$) and all of them have two fixed points through which the Universe evolves at different epochs. On the other hand, for models like matter + CPL DE, one can not, in general, write such an exact evolution equation for $w_{\rm T}$. Moreover, for a sigmoid nature of $w_{\rm T}$, there will be always a maximum in $\frac{dw_{\rm T}}{dz}$ (For details, see Appendix \ref{appen}). For CPL DE model, this is not obvious.

Since the logistic description characterizes the cosmic evolution through an RG-like flow of $w_{\rm T}$, the most natural diagnostic for deviations from the concordance model is the jerk parameter, as already evident from Eq.~\eqref{jerkwt}. Consequently, any observationally significant deviation from $j=1$ would constitute a clear signature of cosmological evolution beyond the standard $\Lambda$CDM scenario.\\

\section{Datasets and Methodology}
We constrain the model using the following observational datasets: 13 correlated BAO distance measurements from the DESI-DR2 release taken from Table IV of~\cite{DESI:2025zgx} which we refer to as DESI-DR2; the latest SNIa sample from the five-year DES supernova program~\cite{DES:2025sig} which is denoted as DES;
constraints from the CMB distance priors which consists of the CMB shift parameter, the acoustic angular scale, and the present physical baryon density $\omega_b\equiv\Omega_b h^2$, where $h=H_0/(100\,{\rm km\,s^{-1}\,Mpc^{-1}})$ providing a compact summary of the CMB information relevant for background cosmology~\cite{Wang:2007mza,Bansal:2025ipo}. 

We specifically use the combined likelihood from \emph{Planck}~\cite{Planck:2019nip} and Atacama Cosmology Telescope~\cite{ACT:2023kun} and follow the method discussed in~\cite{Bansal:2025ipo}. We take into account the radiation contribution accordingly while incorporating the CMB information. In the absence of CMB distance priors, we impose a Gaussian prior on the sound horizon at the drag epoch, $r_d = 147.09 \pm 0.27 \ {\rm Mpc}$ consistent with measurements from the CMB. Throughout the analysis this prior is denoted by $r_d^{\rm plc}$. Our use of CMB distance priors follows standard practice in background level analyses. Since our model modifies only the late-time expansion history and does not introduce new physics at recombination, we assume that the impact on perturbations is subdominant.

To constrain the model parameters we perform a Bayesian Markov Chain Monte Carlo (MCMC) analysis using the \texttt{emcee} sampler~\cite{Foreman_Mackey_2013}.  
Post-processing and visualization of the chains are carried out using \texttt{GetDist}~\cite{Lewis:2019xzd}. To evaluate the goodness of fit we compute the minimum $\chi^2_\text{min}$ corresponding to the maximum a posteriori (MAP) solution using the \texttt{iminuit} minimization algorithm~\cite{James:1975dr}. In addition, Bayesian evidence~\cite{Trotta:2017wnx, Trotta:2008qt} is calculated using the \texttt{MultiNest} algorithm~\cite{Feroz:2008xx} in order to perform model comparison. The priors used in our analysis are mentioned in Table~\ref{priors}.\\

\begin{table}
\begin{center}
\renewcommand{\arraystretch}{1.3}
\setlength{\tabcolsep}{15pt}
\resizebox{0.45\textwidth}{!}{
\begin{tabular}{lccc}
\hline\hline

\multicolumn{4}{c}{\textbf{DESI-DR2+DES+$r_d^{\rm plc}$}} \\
\hline
{\bf Parameter} & {\bf $\Lambda$CDM} & {\bf Logistic} & {\bf CPL} \\
\hline

$\Omega_{m0}$ & $(0,\,0.7)$ & -- & $(0,\,0.7)$ \\
$A$           & --          & $(0,\,3)$ & -- \\
$B$           & --          & $(0,\,3)$ & -- \\
$w_0$         & --          & --        & $[-3,\,1]$ \\
$w_a$         & --          & --        & $[-3,\,2]$ \\

\hline
\multicolumn{4}{c}{\textbf{DESI-DR2+CMB \& DESI-DR2+DES+CMB}} \\
\hline
{\bf Parameter} & {\bf $\Lambda$CDM} & {\bf Logistic} & {\bf CPL} \\
\hline

$A$   & -- & $(0,\,3)$ & -- \\
$B$   & -- & $(0,\,3)$ & -- \\
$w_0$ & -- & --        & $[-3,\,1]$ \\
$w_a$ & -- & --        & $[-3,\,2]$ \\

\hline\hline
\end{tabular}
}
\end{center}
\caption{Priors on the relevant model parameters. A common prior $h \in (0.1,\,1.2)$ is used for all cases. While including CMB information, the physical baryon and cold dark matter densities are varied as $\omega_b=\Omega_b h^2 \in (0.01,\,0.03)$ and $\omega_c=\Omega_c h^2 \in (0.01,\,0.6)$.}
\label{priors}
\end{table}

\begin{figure}   \hspace{-0.5cm}\includegraphics[width=0.48\textwidth]{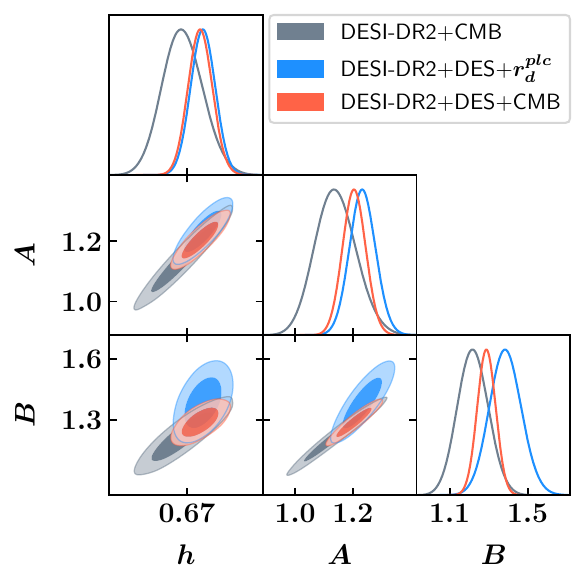}
    \caption{One-dimensional marginalized posterior distributions, 2 dimensional contour plots at 68\% and 95\% C.L. limits for the parameters of the Logistic model using different dataset combinations.}
    \label{fig1}
\end{figure}

\begin{table*}
\begin{center}
\renewcommand{\arraystretch}{1.3}
\setlength{\tabcolsep}{3pt}
\resizebox{0.98\textwidth}{!}{
\begin{tabular}{lccc|ccc}
\hline\hline
Dataset & $h$ & $A$ & $B$ & \multicolumn{3}{c}{$\chi^2_{\rm min}$} \\
\cline{5-7}
 &  &  &  & {\bf Logistic} & {\bf $\Lambda$CDM} & {\bf CPL} \\
\hline

{\bf DESI-DR2+CMB }
& $0.6681\pm0.0091$ 
& $1.141\pm0.071$ 
& $1.220\pm0.078$ 
& 8.95 & 23.44 & 10.33 \\

{\bf DESI-DR2+DES+$r_d^{\rm plc}$ }
& $0.6773\pm0.0055$ 
& $1.234\pm0.045$ 
& $1.386\pm0.082$ 
& 1647.2 & 1654.5 & 1647.2 \\

{\bf DESI-DR2+DES+CMB} 
& $0.6760\pm0.0055$ 
& $1.205\pm0.040$ 
& $1.289\pm0.047$ 
& 1649.2 & 1668.1 & 1653.2 \\

\hline\hline
\end{tabular}}
\end{center}
\caption{Constraints (68\% C.L.) on the Logistic model parameters for different dataset combinations along with comparison of minimum $\chi^2$ values for different models.}
\label{tab1}
\end{table*}

\section{Results}
We present the observational constraints on the Logistic model in Fig.~\ref{fig1} and Table~\ref{tab1}. We carry out the analysis for three combinations of datasets: DESI-DR2+CMB, DESI-DR2+DES+$r_d^{\rm plc}$, and DESI-DR2+DES+CMB. The observational data favor this logistic evolution in all cases, yielding tight constraints on the model parameters. In particular, for the full DESI-DR2+DES+CMB combination we obtain
$A = 1.205 \pm 0.040$ and $B = 1.289 \pm 0.047$ at $68\%$ confidence level (C.L.). Importantly, the Logistic model provides a better fit to the data compared to both $\Lambda$CDM and CPL, as indicated by the corresponding $\chi^2$ values. This improvement is most notable for the DESI-DR2+DES+CMB combination. Although the CPL model also improves the fit compared to $\Lambda$CDM, the Logistic scenario yields further improvement in terms of $\chi^2$. Given that both Logistic and CPL models have equal number of parameters for this combination, this improvement indicates a clear statistical advantage for the Logistic parametrization.

To further quantify the statistical preference among the competing models with the DESI-DR2+DES+CMB combination, we compute several model-selection diagnostics summarized in Table~\ref{tab:model_comparison}. The Logistic model yields a significantly lower Akaike Information Criterion (AIC)~\cite{Akaike1974, Liddle:2007fy} with $\Delta{\rm AIC}=-14.9$ relative to $\Lambda$CDM. The Bayesian Information Criterion (BIC)~\cite{Schwarz1978, Liddle:2007fy} also shows a preference for the Logistic model with $\Delta{\rm BIC}=-4.0$, while the CPL model remains statistically comparable to $\Lambda$CDM under this metric. 
 
We also analyze the model preference using the Bayesian evidence. The resulting evidence difference, $\Delta \ln Z = +1.8$ for the Logistic model relative to $\Lambda$CDM, indicates a positive preference, whereas the CPL model yields $\Delta \ln Z = +1.4$, corresponding to a weaker but still comparable preference~\cite{Trotta:2008qt}. Overall, these results consistently indicate that the Logistic parametrization provides an improved description of the observational data compared to the standard $\Lambda$CDM model, while showing a consistent statistical advantage over the CPL parametrization. 

\begin{figure*}
    \centering
    \includegraphics[width=0.99\linewidth]{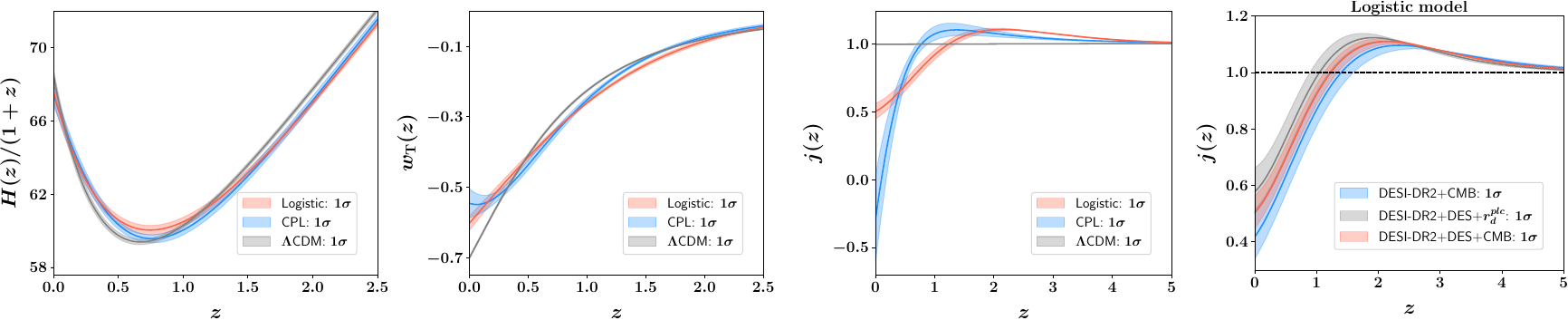}
    \caption{Reconstructed $H(z)/(1+z)$, $w_{\rm T}(z)$, and $j(z)$ for the Logistic, $\Lambda$CDM and CPL models using the constraints obtained from the DESI-DR2+DES+CMB dataset combination along with $j(z)$ for the Logistic model obtained from different dataset combinations.}
    \label{fig2}
\end{figure*}

\begin{table}
\centering
\begin{center}
\renewcommand{\arraystretch}{1.3}
\setlength{\tabcolsep}{10pt}
\resizebox{0.45\textwidth}{!}{
\begin{tabular}{lccc}
\hline\hline
 & {\bf $\Lambda$CDM} & {\bf Logistic} & {\bf CPL} \\
\hline

$k$ & 3 & 5 & 5 \\

$\chi^2_{\rm min}$ & 1668.1 & 1649.2 & 1653.2 \\

AIC & 1674.1 & 1659.2 & 1663.2 \\
$\Delta$AIC & -- & $-14.9$ & $-10.9$ \\

BIC & 1690.4 & 1686.4 & 1690.4 \\
$\Delta$BIC & -- & $-4.0$ & 0.0 \\

$\ln Z$ & $-850.6$ & $-848.8$ & $-849.2$ \\
$\Delta \ln Z$ & -- & $+1.8$ & $+1.4$ \\

\hline\hline
\end{tabular}
}
\end{center}
\caption{Model comparison for $\Lambda$CDM, Logistic, and CPL using DESI-DR2+DES+CMB. Differences are computed relative to $\Lambda$CDM.}
\label{tab:model_comparison}
\end{table}

Having established the observational preference for the Logistic model, we now examine its cosmological implications using the DESI-DR2+DES+CMB dataset combination. The reconstructed evolution of the Hubble parameter, shown in terms of $H(z)/(1+z)$, is presented in Fig.~\ref{fig2} together with the predictions of $\Lambda$CDM and CPL. The qualitative $H(z)$ evolution for all three models is similar, with the Logistic model showing a larger deviation from $\Lambda$CDM around $z\sim 0.7$. Both the Logistic and CPL models exhibit a similar epoch of transition from a decelerating to an accelerating phase, identified by the minimum of the curve, which occurs slightly earlier than in the $\Lambda$CDM model. 

To further illustrate the differences between these models, we also show the evolution of $w_{\rm T}(z)$ in Fig.~\ref{fig2}. At low redshifts, the Logistic model exhibits a noticeably different evolution compared to both the $\Lambda$CDM and CPL predictions (further discussed in the Appendix \ref{appen}). At higher redshifts, all models converge toward $w_{\rm T}\approx0$, corresponding to the expected matter-dominated regime. Interestingly, the behavior of $w_{\rm T}(z)$ at low redshifts for the Logistic model remains quite similar to that obtained from non-parametric reconstructions~\cite{Mukherjee:2025ytj}.

The behavior of $H(z)$ and $w_{\rm T}(z)$ for the three models indicates fundamental differences in their underlying dynamics. An elegant way to quantify this is through the jerk parameter which is exactly $j=1$ for a $\Lambda$CDM universe. Consequently, any significant deviation from $j=1$ signals a decisive departure from the standard paradigm. The evolution of the jerk parameter $j(z)$ for the Logistic and CPL models, together with $j=1$ from $\Lambda$CDM, is shown in Fig.~\ref{fig2}. Both models exhibit clear deviations from the $\Lambda$CDM prediction, with the Logistic model displaying a more stable behavior, unlike CPL for which $j$ falls sharply around the present day. The reconstructed values of the jerk parameter at different redshifts are listed in Table~\ref{tab3}. The Logistic model shows large statistically significant deviations from $j=1$ over a wide range of redshifts. In contrast, while the CPL parametrization also departs from the $\Lambda$CDM, it does so with a lower statistical significance.

\begin{table}
\begin{center}
\renewcommand{\arraystretch}{1.8}
\setlength{\tabcolsep}{3pt}
\resizebox{0.45\textwidth}{!}{
\begin{tabular}{l|cc|cc}
\hline
\hline
\bf $z$ &
\multicolumn{2}{c|}{\bf Logistic} &
\multicolumn{2}{c}{\bf CPL} \\

\cline{2-5}

& \bf $j(z)$ & \bf Tension $(\sigma)$ 
& \bf $j(z)$ & \bf Tension $(\sigma)$ \\

\hline

0.0 & $0.506^{+0.052}_{-0.062}$ & $9.5$ & $-0.295\pm0.321$ & $4.03$ \\

0.1 & $0.533^{+0.048}_{-0.057}$ & $9.73$ & $-0.004\pm0.262$ & $3.83$ \\

0.2 & $0.566^{+0.045}_{-0.052}$ & $9.64$ & $0.251\pm0.201$ & $3.73$ \\

0.3 & $0.606^{+0.043}_{-0.049}$ & $9.16$ & $0.465\pm0.146$ & $3.66$ \\

0.4 & $0.650^{+0.042}_{-0.046}$ & $8.33$ & $0.639\pm0.100$ & $3.61$ \\

0.5 & $0.697\pm0.043$ & $7.05$ & $0.777\pm0.067$ & $3.33$ \\

1.0 & $0.922\pm0.035$ & $2.23$ & $1.081^{+0.042}_{-0.057}$ & $1.42$ \\

\hline\hline
\end{tabular}}
\end{center}
\caption{Reconstructed values of $j(z)$ at different redshifts for the Logistic and CPL models using the DESI-DR2+DES+CMB combination. Tension is computed relative to the $\Lambda$CDM prediction $j=1$.}
\label{tab3}
\end{table}

 As evident from Fig.~\ref{fig2}, the DESI-DR2+DES+CMB combination leads to substantially stronger deviations in $j(z)$ compared to the constraints obtained from low-redshift datasets alone. This indicates that the early Universe information encoded in the CMB, when combined with the low-redshift measurements from DESI-DR2 and DES, helps to significantly sharpen the distinction between these different cosmic expansion histories.

\begin{figure}
    \centering
    \includegraphics[width=0.45\textwidth]{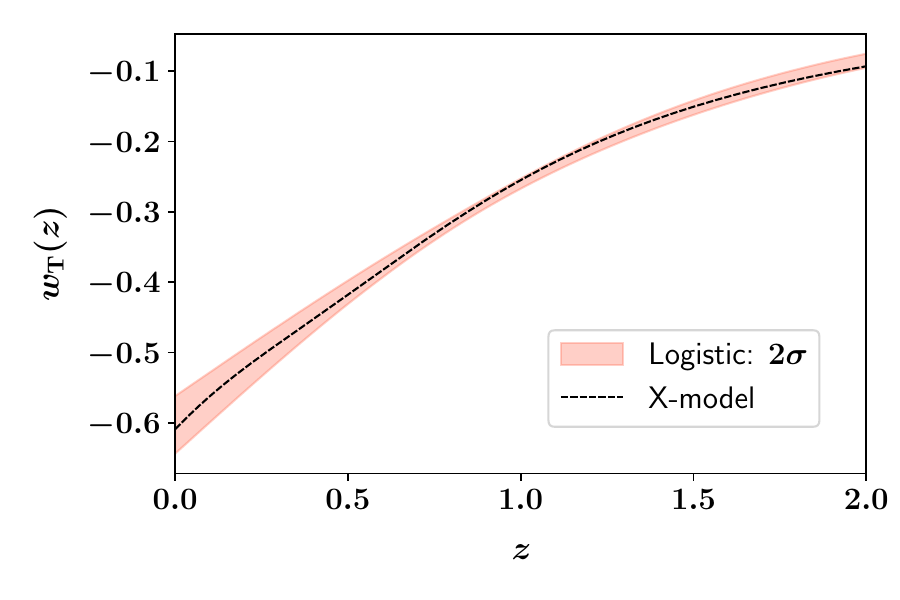}
    \caption{Comparison of $w_\text{T}$ for the Logistic model and a model with an additional $X$-fluid.}
    \label{figwtx}
\end{figure}

While the preceding diagnostics establish a clear observational preference for our phenomenological framework, it is crucial to explore its physical origin. The resulting evolution of $w_{\rm T}$ can naturally arise from an underlying multi-component dark sector, providing a possible theoretical foundation for our results. Recently, to resolve the phantom-crossing behavior observed in single-component DE models, several studies~\cite{Caldwell:2025inn, Liu:2025bss} have proposed introducing an extra fluid component (which we denote as the $X$-fluid) alongside the cosmological constant. The equation of state for this $X$-fluid transitions from $w_\text{X}\approx -1$ at early times to $w_\text{X}\approx 0$ at recent epochs. Such dynamics naturally emerge in various physically motivated settings, including axion-like fields~\cite{Preskill:1982cy, Abbott:1982af, Dine:1982ah, Carroll:1998zi, Shajib:2025tpd}, generalized Chaplygin gas~\cite{Sen:2005sk} models, etc. As a representative example, we consider
\begin{equation}
   w_\text{X}= -\frac{1}{2} \left[1 + \tanh\left(\frac{z-z_0}{\Delta}\right)\right].
\end{equation}
Figure~\ref{figwtx} illustrates that the resulting  $w_\text{T}$ for this multi-component model, with reference values $\Omega_{m0} = 0.2974$, $\Omega_{\Lambda} = 0.570$, $z_0 = 0.198$, and $\Delta = 0.451$, shows excellent agreement with that of the Logistic model for DESI-DR2+DES+CMB combination. A detailed study of this physical correspondence is beyond the scope of this work and will be pursued elsewhere.\\

\section{Conclusions}
To summarize, we have presented a logistic RG-flow-inspired parametrization of the total equation of state $w_{\rm T}(z)$, providing a simple and physically motivated description of the transition from matter domination to late-time acceleration. By focusing directly on $w_{\rm T}(z)$~\cite{Caldwell:2025inn,Liu:2025bss}, this framework avoids assumptions about the underlying composition of DE and offers a complementary perspective to conventional parametrizations.

Using DESI-DR2 BAO, DES SNIa data, and CMB distance priors, we find that the Logistic model provides an improved description of the expansion history relative to $\Lambda$CDM and exhibits a consistent statistical advantage over the widely used CPL parametrization. Furthermore, the derived jerk parameter indicates statistically significant deviations from the $\Lambda$CDM expectation at lower redshifts. We emphasize that these deviations are parametrization-dependent and should not be interpreted as model-independent evidence against $\Lambda$CDM, but rather as a demonstration that current data show a preference for alternative descriptions.

As a concluding remark, the Logistic framework offers a minimal yet elegant phenomenological description of cosmic evolution as a flow between dynamical fixed points. Our results suggest that this simple flow-inspired approach can capture features of the late-time expansion history that are not easily accommodated by standard minimal models. Future high-precision surveys such as LSST, Euclid, and SKA will provide stringent tests of these features and help assess whether flow-based descriptions capture genuine physical aspects of late-time cosmic expansion.

\section*{Acknowledgments}
SGC acknowledges funding from the Anusandhan National Research Foundation (ANRF), Govt. of India, under the National Post-Doctoral Fellowship (File no. PDF/2023/002066). AAS acknowledges the funding from ANRF, Govt. of India, under the research grant no. CRG/2023/003984. This article/publication is based upon work from COST Action CA21136- ``Addressing observational tensions in cosmology with systematics and fundamental physics (CosmoVerse)'', supported by COST (European Cooperation in Science and Technology). We thank Purba Mukherjee for helpful discussions and assistance with the implementation of the CMB compressed likelihood.

\bibliography{ref}
\appendix

\section{Theoretical motivation for a logistic evolution of the cosmic equation of state $w_{\rm{T}}$}\label{appen}
\begin{figure*}[t]
    \centering
    \includegraphics[width=0.48\textwidth]{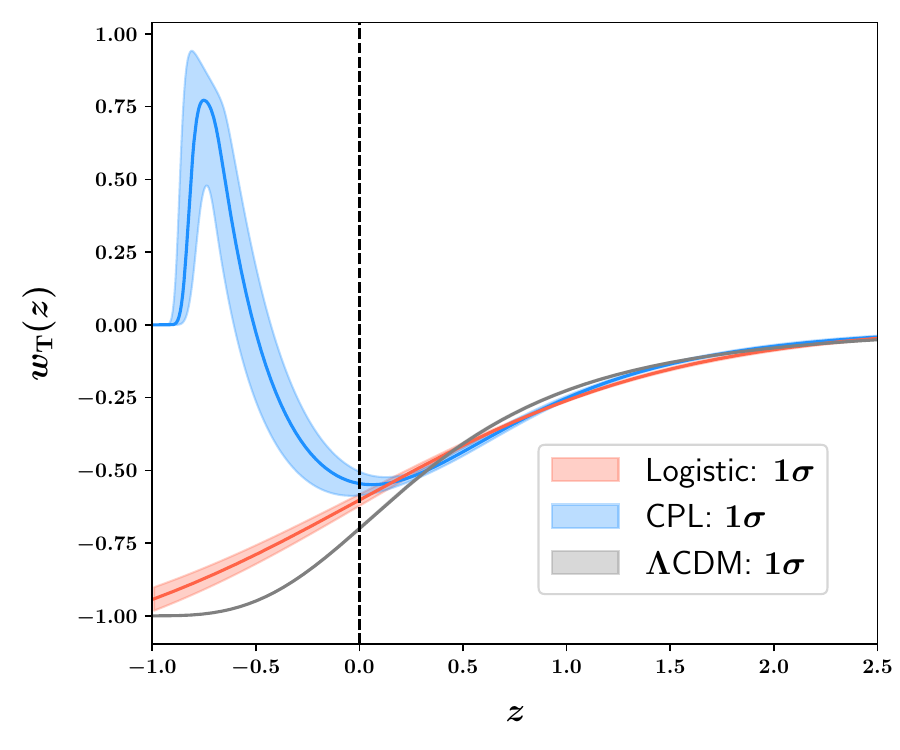}
    \includegraphics[width=0.48\textwidth]{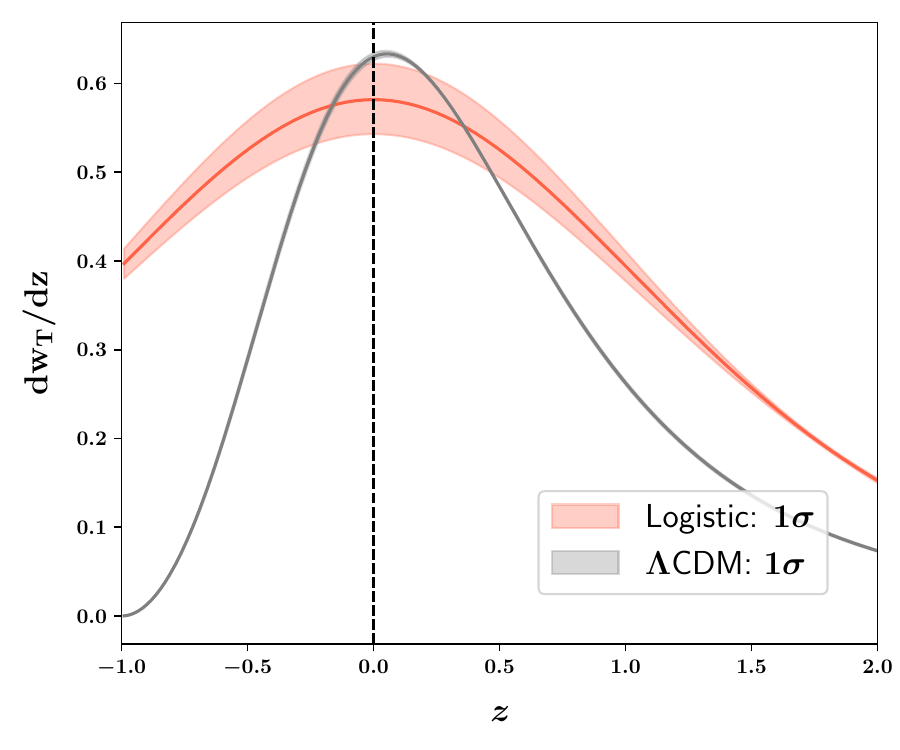}
    \caption{Evolution of $w_{\rm T}$ (left panel) and $\frac{{\rm d}w_{\rm T}}{{\rm d}z}$ (right panel) extrapolated into the future for the considered models.}
    \label{wtfuture}
\end{figure*}
The total equation of state is related to the deceleration parameter through
\begin{equation}
q(z) = -\frac{\ddot a}{aH^2} = \frac{1}{2}(1+3w_{\rm T}),
\end{equation}
so that specifying $w_{\rm T}(z)$ completely determines the dynamics of the background expansion history.
For comparison, $w_\text{T}$ in the $\Lambda$CDM and CPL models is given by
\begin{equation}\label{wtlcdm}
w_{\rm T}^{\Lambda}=
\frac{\Omega_{m0}-1}{\Omega_{m0}(1+z)^3+(1-\Omega_{m0})},
\end{equation}
and
\begin{equation}
w_{\rm T}^{\text{CPL}} =
\frac{
(1-\Omega_{m0})\left(w_0+\frac{w_a z}{1+z}\right)f(z)
}{
\Omega_{m0}(1+z)^3+(1-\Omega_{m0})f(z)
},
\end{equation}
where
\begin{equation}
f(z)=(1+z)^{3(1+w_0+w_a)}
\exp\!\left(-\frac{3w_a z}{1+z}\right).
\end{equation}

The key difference in the behavior of $w_\text{T}$ for $\Lambda$CDM and the Logistic model arises from the choice of flow variable in the logistic evolution:
\begin{equation}\label{dwtlcdm1}
\frac{dw_\text{T}^{\Lambda}}{d \log a}=3\,w_\text{T}^{\Lambda}(1+w_\text{T}^{\Lambda}),
\end{equation}
and
\begin{equation}\label{dwtlogis}
\frac{dw_{\rm T}}{dz} = -\frac{B  w_{\rm T}\left(A + w_{\rm T}\right)}{A}.
\end{equation}

Such a structure cannot be realized for the CPL parametrization for nontrivial values of $w_0$ and $w_a$. From this perspective, $\Lambda$CDM and the Logistic model belong to the same class of logistic-type evolution, motivating the latter as a natural extension, though not embedding $\Lambda$CDM as a limiting case. In both models, $w_\text{T}$ can be interpreted as a flow between cosmological fixed points corresponding to matter domination and late-time acceleration.

The left panel of Fig.~\ref{wtfuture} shows that the transition from matter domination to acceleration is sharper in $\Lambda$CDM than in the Logistic model yet both models remain well-behaved when extrapolated into the future. In contrast, the CPL model exhibits a completely different qualitative behavior, reflecting an artifact of its parametrization.

Equations~\eqref{dwtlcdm1} and \eqref{dwtlogis} imply that $\frac{{\rm d}w_\text{T}}{{\rm d}z}$ attains a maximum at $w_\text{T}=-\frac{2}{3}$ for $\Lambda$CDM (independent of $\Omega_{m0}$) and at $w_\text{T}=-\frac{A}{2}$ for the Logistic model. The latter occurs at $z=0$, while for $\Lambda$CDM the corresponding redshift depends on $\Omega_{m0}$. This is also evident in the right panel of Fig.~\ref{wtfuture}, where both models show qualitatively similar behavior. However, the significantly flatter evolution of the Logistic model leads to the large deviations observed in the jerk parameter.

\end{document}